\documentclass[runningheads]{llncs}

\usepackage{tikz}
\usepackage{amsmath}
\usepackage{booktabs}
\usepackage{array,multirow,graphicx}
\usepackage{afterpage}
\usepackage{algpseudocode}
\usepackage{algorithm}
\usepackage{xcolor}
\usepackage{hyperref}
\usepackage{enumitem,kantlipsum}

\usepackage{caption}
\usepackage{url}

\usepackage{breakurl}

\usepackage{graphicx}
\usepackage{hyperref}
\usepackage{filecontents}
\newcommand {\toolname }{MUD-Visualizer}
\newcommand {\mudfile }{MUD-File}
\newcommand{\acetree}{ACE Tree}
\newcommand{\acepruning}{ACE Pruning}
\newcommand{\acemerging}{ACE Merging}
\begin{document}
\title{On the Analysis of MUD-Files' Interactions, Conflicts, and Configuration Requirements Before Deployment}
\titlerunning{Analyzing MUD-Files Before Deployment}

% \author{Authors removed for anonymity
% }

\author{Vafa Andalibi\inst{1}\orcidID{0000-0003-1517-3185} \and
Eliot Lear\inst{2}\orcidID{0000-0003-2724-0293} \and
DongInn Kim\inst{1}\orcidID{0000-0001-6331-1410}
\and
L. Jean Camp\inst{1}\orcidID{0000-0001-8731-7884}
}

\authorrunning{Andalibi et al.}

\institute{Indiana University, Bloomington, IN 47408, USA 
\email{\{vafandal,dikim,ljcamp\}@indiana.edu}\\
\and
Cisco Systems, Zurich, Switzerland\\
\email{lear@cisco.com}}

\maketitle              % typeset the header of the 

\begin{abstract}
Manufacturer Usage Description (MUD) is an Internet Engineering Task Force (IETF) standard designed to protect IoT devices and networks by creating an out-of-the-box access control list for an IoT device. %The protocol defines a conceptually straight-forward method to implement an isolation-based defensive mechanism based on the rules that are introduced by manufacturer of the device. However, in practice the
  Access control list of each device is defined in its \mudfile{} and 
  may contain possibly hundreds of access control rules.
  As a result, reading and validating these files is a challenge; and determining how multiple IoT devices interact is difficult for the developer and infeasible for the consumer. 
  To address this we introduce the 
 \toolname{} to provide a visualization of any number of \mudfile{}s. \toolname{} is designed to enable developers to produce correct \mudfile{}s by providing format correction, integrating them with other \mudfile{}s, and identifying conflicts through visualization. \toolname{} is scalable and its core task is to merge and illustrate ACEs for multiple devices; both within and beyond the local area network. \toolname{} is made publicly available and can be found in GitHub. 

\end{abstract}
\section{Introduction}

The Internet of Things (IoT) has diffused across the globe and the estimates of IoT devices in the home range from billions to tens of billions. Yet, security has lagged \cite{iot_analytics_2018}. 
 The security of IoT devices is such that they are used to participate in DDoS attacks \cite{kolias2017ddos}, are vulnerable to ransomware \cite{yaqoob2017rise}, and enable information exfiltration from within homes \cite{d2016data}. Media reports of abusive strangers engaging with families through IoT devices are not uncommon, e.g. \cite{ring_doorbell_abcnews}. Given the complexity of IoT devices, the lack of technical support, the level of technical expertise in the home,  and the complexity of access control, how can these devices be managed?

% - MUD as an isolation based solution
Manufacturer Usage Description (MUD) is an Internet Engineering Task Force (IETF) standard created in response to the requirements for access control and device isolation for IoT devices \cite{rfc8520}. It addresses multiple challenges regarding IoT security by relying on manufacturers providing an Access Control List (ACL) that identifies services and addresses for those services. The goal is to isolate devices; particularly those that cannot be relied upon to provide their own protection. Unlike more traditional verification approaches, MUD can work with devices that have highly limited processing power. In addition, rather than a single entity creating policy, each manufacturer creates the access control that defines the situation for their own devices. 
% However, the verification of a policy in a dynamic network is known to be difficult. 
% Similarly, writing correct \mudfile{}s which includes the access control information will also be a challenge. 
The second goal of the MUD standard is to provide an identifier so that updates to devices can be implemented only from authenticated and authorized sources. With such functionality, errors in device configurations can be mitigated. 
% One of the goals of the the \toolname{} is to decrease such errors. 

We have chosen to focus on MUD because, in addition to being an IETF standard, MUD is also a core component of the \textit{National Institute of Standards and Technology (NIST) security for IoT Initiatives}, particularly the thrust focused on stopping DDoS \cite{dodson2019securing}.  MUD can defend IoT devices in a home from other compromised ones in the household and on the network, with a specific goal of blocking the access of compromised devices to command and control channels. 

One of the core components of the MUD is the \mudfile{} which is essentially an access control statement. The \mudfile~ enumerates the allowed (or specifically disallowed) services and sources for these services. In the MUD standard, it is defined as ``a file containing YANG-based JSON that describes a Thing and associated suggested specific network behavior'' \cite{rfc8520}. \mudfile{}s could possibly be long and complex, making their reading, reviewing, and modification a laborious task if performed manually.

In this paper, we present \toolname{} that addresses this issue. \toolname~ provides 1) protocol checking to avoid formatting errors in the \mudfile, 2) optimization of the \mudfile{} which identifies internal inconsistencies and inefficiencies, and 3) visualization of the commands in \mudfile{}s. The first of these prevents coding errors. The second prevents logic errors. The third enables manufacturers and sysadmins to review, validate, and modify the \mudfile{}s prior deployment. The focus on coding, logic, and contextual errors aligns with the sources of most vulnerabilities~\cite{landwehr1994taxonomy}.

% ------------------------------------------------------------
\section{Manufacturer Usage Description (MUD)} \label{sec:mud}
% ------------------------------------------------------------

Understanding the importance of the \mudfile{}s requires some understanding of the MUD standard. For the readers not familiar with the workflow of MUD a brief summary of MUD workflow and its abstractions is presented in this section. Those familiar with MUD may want to continue to Section~\ref{sec:motivation}.

\subsection{Components and Workflow}

\begin{figure}[h!]
    \begin{center}
    \includegraphics[width=0.44\linewidth]{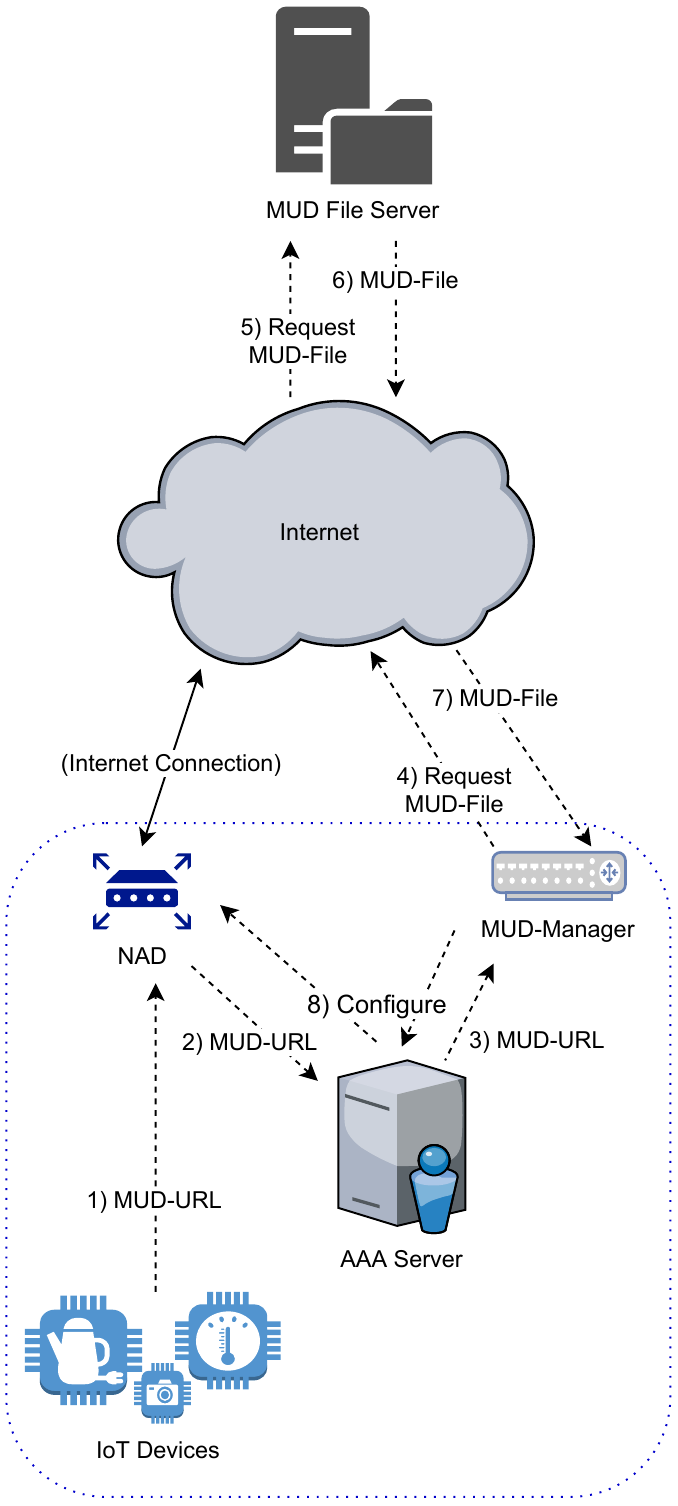}
    \caption{MUD workflow in a LAN: the blue dotted line indicates the boundary of the LAN }
    \label{fig:mud-workflow}
    \end{center}
\end{figure}

An implementation of MUD has six main components as presented in Fig.  \ref{fig:mud-workflow}: 
\begin{enumerate}
% [wide, labelwidth=!, labelindent=0pt]
    \item \textit{\mudfile{}:} is a YANG-based JSON file (RFC 7951) created, signed with a public key signature, and distributed by the manufacturer that describes the expected network behavior of the device.
    \item \textit{MUD file server:} on which \mudfile{} is hosted and the location of the file is embedded as a uniform resource locator called MUD-URI.
    \item \textit{MUD-URI}: This is used to locate and download the \mudfile{} to the local network.
    \item \textit{AAA Server}: The Authentication, Authorization, and Accounting (AAA) server enforces the traffic rules on the devices in the network. This server can be either an independent server or a built-in component in the \textit{Network Access Device (NAD)}.
    \item \textit{Network Access Device (NAD)}: acts as a router in the network and is usually equipped with an internal Firewall component which is used by the MUD-Manager via AAA server to control the traffic and enforce rules.
    \item \textit{MUD-Manager}: is the core of MUD architecture and is responsible for receiving the MUD-URI from the devices, retrieving the \mudfile{} from the MUD file server, and communicating the \mudfile{} rules to the AAA server. 
\end{enumerate}

The MUD workflow illustrated in Fig.  \ref{fig:mud-workflow} begins with the IoT device authenticating with X.509 certificate (although DHCP and LLDP are also available as means of authentication) transmitting the MUD-URI embedded in the device to the NAD. MUD-Manager will then receive the MUD-URI, validate the signature, retrieve the \mudfile{} from the manufacture's MUD file server and then enforce the access control rules of the MUD file to the network via AAA server.

\subsection{\mudfile{} ACL Abstractions}
\label{sec:abstractions}

The content of and ACLs in the \mudfile{}s are instantiated as firewall rules by the MUD-Manager. The ability to implement such rules enables a range of policies. 
%An open research question is how to enable home owners to interact with these policies  
Based on the MUD specifications, the contents of the \mudfile{} may include a range of default policies and defined types. There are seven approaches to define the behavior and constraints of a device in a MUD instance. These include constraints or identification of domain names, manufacturers, device class, device models, and the local context of the device. These extensions to the IETF-ACL were all addressed in our implementation. Each extension has the potential to simplify the use of MUD for the manufacturer and adopter. Yet the existence of all these options also drives the need for \toolname.

A useful abstraction for cloud access is the \textbf{domain-name}. Of course, domain names are a global Internet namespace; which is not always well-suited for a specific, geographically located IoT device. 

A second abstraction that is defined in the MUD specification is that of the \textbf{local-networks}. With local-networks abstraction, a node will be matched against the nodes in the same network. This is particularly useful for designs where there is an IoT hub in the house that is compliant with the local devices.

Another constraining abstraction is that of the \textbf{manufacturer}. In this case, the Hostname of the target node would match against the authority component, e.g., domain name, or MUD URL of another node. This constrains the reach of a device to other devices from particular manufacturers. 
 
Beyond the basic domain name for every IoT device, MUD provides the indicator of \textbf{same-manufacturer}. In this abstraction, multiple devices, e.g., a house where its lightbulbs or outlets are from the same creator, can communicate only with each other. In operational terms, when devices use this extension the authoritative component will be checked against that of another node.
 
More than one extension can be used in the same file. Thus there is also an option that identifies a device requiring contextual configuration. The \textbf{controller} extension identifies the care where the network administrator needs to assign the target devices to a particular class. This may be particularly useful when a single device embeds multiple services, and access control depends on these services. For example, a doorbell that offers wireless installation may also offer remote control for the owner, access to a home security firm, face recognition for local control, real-time video activity monitoring, social interactions (e.g., sharing moments), caregiving, and other services which impinge delegating access, network connections, data types, and port numbers.

With \textbf{my-controller} abstraction, the device will leverage its MUD-URL and will signal the MUD manager to use whatever mapping it has for this MUD URL to a particular group of hosts that would be used to manage or control this class of device. This mode requires a local decision-maker in the loop, whether that is a human or a process. With this option, the node initiates communication with the MUD-manager with a request that the MUD-Manager assign this node to a class.

Classes of devices and devices from the same manufacturer may be inadequately constraining. The seventh requirement is that model, manufacturer, and class must all match, defined as \textbf{model}. The potential for conflicts with this set of seven defined options argues for the importance of the visualizing tool.

%-------------------------------------------------------------------------------
\section{Motivation}\label{sec:motivation} % threat model
%-------------------------------------------------------------------------------

As described above, MUD's role in enforcing the Principle of Least Privilege (PoLP) is to limit the devices' reachability to a bare minimum by leveraging the manufacturers' knowledge about their devices. For a given configuration the ACLs should be validated (if not defined) manually to ensure PoLP, which makes this process prone to human error \cite{maxion2005improving,wang2019accessible}. Errors in the ACLs defined by the manufacturer will result in unwanted access control privilege escalation which poses a security risk to the network. 

For instance, consider a MUD-compliant smart slow cooker device that is only allowed to communicate with the manufacturer server and the manufacturer`s site over the Internet. In addition, the slow cooker comes with a mobile application that allows controlling its functionality both online (via the Internet) and offline (locally). When used offline, the slow cooker will only use port 1300 for communication. Alice is a sysadmin in an enterprise network with more than 100 types of MUD-compliant devices. Bob decides to add the smart slow cooker to the kitchen of the office as a collegial act. A new \mudfile{} appears in Alice`s MUD-Manager. To understand all the possible connections, Alice would be required to have an encyclopedic awareness of each \mudfile{} in the network, and be able to read these files while understanding the interactions.

Given the reality of the enterprise, it is almost impossible to prevent employees from bringing in personal fitness trackers, slow cookers, rice cookers, microwaves, or space heaters all of which can be Internet-connected devices. Because of the number of devices and corresponding ACLs in the network, it's arguably beyond human cognition for Alice to identify the possible unnecessary communication between the slow cooker and the smart bulb which puts the network at risk with only text files \cite{wang2019accessible}. Although there exist other tools which have the potential to identify the connection between the light bulbs and slow cookers or between the slow cooker and the attacker, \toolname{} is one step ahead and tries to prevent these before they occur. 

Imagine instead Alice has the \toolname. When each device is added, \toolname{} can help her to implement informed threat modeling and flag unwanted communications. Using \toolname{} Alice could easily identify the connection and isolate the slow cooker from the internal system by adding an additional rule. 

Beyond the aforementioned example, there are several points made in previous research studies indicating the importance of different aspects of a tool like \toolname{}. We mention the most important of such works and corresponding points associated with \toolname{} in the next section.

%-------------------------------------------------------------------------------
\section{Related Work}
%-------------------------------------------------------------------------------
In this section, we present related work that fits in one of the following categories: studies that focus on MUD research and tools, as well as the researches that focus on the importance of Human Computer Interaction (HCI) in mitigating human errors in regards to access control.  Usable access control is a significant challenge, where even the relative responsibilities of the platform, the final user, and the developer are contested, e.g.,~\cite{roesner2012user,tahaei2021developers}.

The tool described here is developed as a complement to the MUD \cite{rfc8520}. At the time of writing this paper, there are four main projects that implement the core of a MUD instantiation: the Cisco MUD-Manager \cite{Weis2018}, the Open Source MUD Manager \cite{Yeich2019}, the MUD implementation of NIST\cite{Ranganathan2018}, and CableLabs Micronets \cite{Pratt2019}. NIST has a special publication that thoroughly describes four different builds of MUD based on the above-mentioned implementations for mitigating network-based attacks \cite{dodson2019securing}. 

Besides the MUD-Manager, Cisco also offers the \textit{mudmaker}\footnote{https://www.mudmaker.org} which can be used for creating \mudfile{}s. We used \textit{mudmaker} to generate a comprehensive \mudfile{} for the tests reported in section \ref{sec:results}. 
MUD Pretty Printer \cite{lear2020mudpp} is another tool developed to summarize ACL information on the MUD files. However, it does not provide any User Interface (UI) or visualization. With regard to \mudfile{} modification, Cisco has a recent patent that discusses techniques for providing secure modification of \mudfile{}s based on the device applications \cite{lear2019secure}. 

Regarding MUD deployment, there are some studies that focus on the effectiveness of MUD against DoS and DDoS attacks \cite{hamza2019detecting,polk2017project,andalibi2019throwing,schutijser2018towards}. Afek et al. \cite{afek2019nfv} proposed an ISP-level system architecture that enforces the ACL upstream at the provider network to protect the IoT at scale. Additionally, they extended their MUD-interoperable architecture to support peer-to-peer protocols. 

With regard to combining Software Defined Networking (SDN) and MUD, the authors in \cite{ranganathanimplementing} present a scalable implementation of the MUD standard on OpenFlow-enabled SDN switches. Hamza et al. in \cite{hamza2018combining} attempted to create flow rules based on MUD policies so that they can be enforced using SDN. Matheu et al. \cite{matheu2020security} employed SDN technique to make MUD model more flexible to support additional aspects such as data privacy, channel protection, and resource authorization. In another work, an SDN-based architecture was proposed to make the process of obtaining and enforcement of MUD policies secure \cite{garcia2019enforcing}. In a proposed expansion to MUD Matth{\'\i}asso presented generating contracts and their local evaluation, \cite{matthiasson2020iot}; this is complimentary to \toolname{} as it assumes the existence of non-conflicting MUD files.

Some researchers focused on helping the manufacturers in the process of creating \mudfile{}s.
In \cite{hamza2018clear} Hamza et al. described \textit{MUDgee} which uses the traffic of an IoT device to generate a MUD profile for that IoT device. Beyond the \mudfile{}s generated from \textit{mudmaker}, we also performed some dry-run tests with the profiles provided by MUDgee project\footnote{https://iotanalytics.unsw.edu.au/mudprofiles}. NIST also has an ongoing open-source project in their GitHub organization entitled \textit{MUD-PD} \cite{Watrobski2018} which is similarly targeted at profiling IoT devices in order to leverage the MUD architecture.

Feraudo et al. in their systematization of knowledge about MUD identified two challenges with \mudfile{}s that can be mitigated with the use of the \toolname{}  \cite{feraudo2020sok}. The first of these is inconsistent implementations; \toolname{} can be used on the underlying files to easily determine developer intent, thus simplifying differences in results. The other is of course consistent generation of \mudfile{}s.  In another survey paper, Mazhar et al. \cite{mazhar2021role} detailed the role of the MUD in the IoT ecosystem, including the implementation, its role in IoT security, and its integration in different security frameworks. They also review the benefits of MUD to the industrial IoT, telecommunication networks, smart home, Fog and Edge computing, and mobile application. \toolname{} can facilitate the deployment of MUD in all of these applications. 

In a similar area of research related to configuration verification, Prabhu et al. \cite{prabhu2020plankton} presented Plankton which is purposed for network configuration verification.  In another study, Fogel et al. \cite{fogel2015general} proposed a high-fidelity declarative model of low-level network configurations and implemented it as a tool called Batfish. Fayaz et al. \cite{fayaz2016efficient} implemented ERA which can be used for bug detection in reachability policies. With regard to routing, ARC \cite{gember2016fast} finds the possible impact of routing protocols on the network’s data plan by abstracting their mechanics. Beckett et al. in \cite{beckett2017general} presented Minesweeper which can be used to ensure a wide range of intended properties, e.g., isolation among nodes, in the network. Unlike \toolname{} none of these offer visualizations. 

The earliest work in optimizing ACL interaction using HCI principles was by Maxion and Reeder \cite{maxion2005improving}. They examine the Windows XP file-permissions and found that the visualizer tripled the rate of assigned task completion, and reduced errors in those completed tasks by up to 94\%. This illustrates the importance of visualization and interaction design in access control. 

Similar to \toolname{} Vaniea et al. our work is grounded in the recognition of the difficulty of translating policy rules into access control rules~\cite{vaniea2008evaluating}. We integrated their recommendations into our design, in particular, we integrate visual feedback while the developer is drafting the \mudfile. SPARKLE~\cite{vaniea2008evaluating} followed the common visualization process of focusing on a presentation of data in a table, which is the visualization approach most commonly used. For example, Reeder et al. \cite{reeder2008expandable} developed an interactive matrix visualization, the Expandable Grid, to enable improved file permissions in Windows XP. This ACL visualization informed the design of \toolname{} particularly in terms of understanding challenges to ease of cognition. In comparison, our approach provides a visualization based on flows more similar to the graph visualization approach by Kolomeets et al~\cite{KolomeetsAccessControlGraph}; rather than asking developers to read rows or columns.

Salim et al. took a different view examining access control as a case of decision-making under uncertainty \cite{salim2011approach}. They provided a formal method to quantify how much uncertainty is inherent in the Role-Based Access Control (RBAC), one that illustrates the level of complexity required to provide reliably correct access in an organization.

Xu and colleagues investigated the role of uncertainty in access control decisions \cite{xu2017system}. They implemented qualitative investigations into how systems administrators resolve access control conflicts, as these human errors are a known source of security vulnerabilities. Their fundamental finding was that a lack of feedback forced administrations into a trail and error mode. \toolname{} provides real-time visual feedback about changes in access control. This verifies that a need for a high-level view providing information about access requirements and settings in a network. The complexity they documented may be far greater with the expansion of IoT.

Smetters et al. \cite{smetters2009users} in their study found that limitations in the UI would lead to the reluctance to change the access control settings. This finding applies to MUD deployment as well; it would be simply difficult and time-consuming for system administrators to manually evaluate the interaction between tens of types of \mudfile{} associated with their IoT device. We believe \toolname{}'s UI is significantly easier to use compared to manual analysis and we are going to thoroughly evaluate and show this in our future work. 

Besides the lack of UI in the manual analysis of \mudfile{}, the other issue with manual analysis is processing errors. Liginlal et al. \cite{liginlal2009significant} focused on the importance of the analysis errors. They found that mistakes in the information processing stage constitute the most cases of human error-related privacy breach incidents, confirming the importance of \toolname{} in \mudfile{} interaction analysis.

Another source of user errors is called goal errors, i.e. the failures of users to understand what to do. The main source of goal errors is found to be poor information representation in the interface. A study of highly skilled programmers found that even these participants struggled with access control~\cite{reeder2005user}. This indicates the importance of information representation of \toolname{} compared to other text-based tools like MUD Pretty Printer \cite{lear2020mudpp}.

From another perspective, the issues with conflicts in \mudfile{}s are comparable to the challenges in the SDN flow information base (FLIB). None of the aforementioned studies address the verification of \mudfile{}s. However, previous work on SDN verification and human subjects research on access control informed the design of the \toolname.

The only work that attempts to help the IoT manufacturers and adopters of these devices in process of preparing or deploying MUD profiles is \cite{hamza2019verifying}. Their work focuses on different aspects compared to this study in two ways. First, similar to \cite{hamza2018clear,Watrobski2018}, they focus on automatic generation of MUD profile based on network traffic. Second, their tool validates the consistency and compatibility of the generated profiles with organizational policies. Because those project generate \mudfile{}s, code-checking is less of a challenge. Conversely since these products create \mudfile{}s automatically, logic errors can still be embedded. Their work does not have a visualization or usability component. \toolname{} is a complement to the projects which automatically generate \mudfile{}s.

To our knowledge, this is the only product targeted at the developers or sysadmins seeking to define or validate a \mudfile{} for a product to be deployed. \toolname{} is also unique in that it validates interactions and identifies possible conflicts prior to deployment (for the manufacturer) or at the time of deployment (for the user). 

% in future work, we now move to the usability literature
% here we focus on validation

%The design of the visualization is also motivated by the work on efficacy and usability of access control systems. 
% maxioms paper
% \cite{maxion2005improving}

% any paper that references his paper and is in NDSI or NDSS

% maybe Mez's orginal paper
% \cite{metz1999aaa}

%-------------------------------------------------------------------------------
\section{Methods}
\label{sec:methods}
%------------------------------------------------------------------------------
Recall from Section~\ref{sec:mud} that the \mudfile{} of each IoT device typically contains access controls in the form of a white list. Each list entry contains information about one or more protocols and often a corresponding identifier, e.g., a domain name accessible only via SSH. The white list provides the identifier appropriate for the network layers of the protocol. Entries in this \mudfile~list are called the Access Control Entries (ACEs).  

The first task of \toolname{} is to determine how the ACE information of different devices interact. We call this process \textit{\acemerging}. When we merge a set of ACEs of two devices, it is often possible that duplicates appear in the final list of merged protocol information. We address this issue by pruning the protocol stacks that are a subset of more generic protocol stacks. We call this process \textit{\acepruning}. Both of these procedures are described in the following subsections.

\begin{algorithm}[t!]
\caption{Merging Two Protocol Stacks}
\label{alg:protocol_merging}
\begin{algorithmic}[1]
\State initialize empty protocols stack $PS_{out}$
\Procedure{MergeProtocolStacks}{$PS_{src}$, $PS_{dst}$}
\For{each layer $l$ in protocol stack}
    \For{each protocol $P_l$ in layer $l$ }
        \If{$P_{l_{src}} \subseteq P_{l_{dst}}$}
            \State $PS_{out}\gets PS_{out} + P_{l_{src}} \cap P_{l_{dst}}$ 
        \EndIf
    \EndFor
\EndFor
\If{isFullStack($PS_{out}$)}
    \State return $PS_{out}$
\Else
    \State return
\EndIf

\EndProcedure
\end{algorithmic}
\end{algorithm}

\subsection{\acemerging} \label{sec-acemerging}

When the abstractions of two devices in the network allows them to communicate, e.g., two devices supporting local network connections, their ACEs should be inspected and merged accordingly. Of course, it is possible that even with matching specifications, two devices should only communicate if there is a common factor between their ACEs. Hence, one of the important tasks of the \toolname{} (specifically \textit{MUD-Network} module described in section \ref{sec:implementations}) is to merge and validate the protocols of ACEs. This task is implemented by moving up the protocol stack and check whether the source protocol (sender) is a subset of the destination protocol (receiver) in that layer. If so, the intersection of the protocols is added to the resulted protocol stack. This procedure is implemented in Algorithm \ref{alg:protocol_merging}.

\begin{algorithm}[t!]
\caption{Building the \acetree{}}
\label{alg:protocol_tree}
\begin{algorithmic}[1]
% \SetAlgoLined
\State initialize \texttt{node} as a tree node\;

\For{each \texttt{ACE} in list of \texttt{ACEs}}
    \State initialize \texttt{node} to \texttt{root}\;
    \State {Updateacetree(\texttt{ACE}, \texttt{node}, \texttt{n})}
\EndFor
\Procedure{Updateacetree}{\texttt{ACE}, \texttt{node}, \texttt{n}}
\If{\texttt{node} is \texttt{null}}
    \State return 
\EndIf
\State \texttt{protocols} $\gets$ GetLayerProtocols(\texttt{ACE}, \texttt{n})
\If{Count(\texttt{protocols}) > 1 }
    \State add a wild-card child $W_n$ to \texttt{node}
    \State \texttt{node}  $\gets W_n$
\Else
    \State add a child $C_n$ to \texttt{node}
    \State \texttt{node}  $\gets C_n$
\EndIf
    \State \texttt{ACE} $\gets$ \texttt{ACE[1:]}
    \State Updateacetree(\texttt{ACE}, \texttt{node}, \texttt{n-1})

\EndProcedure
\Procedure{GetLayerProtocols}{\texttt{ACE}, \texttt{n}}
\State initialize \texttt{protocols} as an array
\For{each protocol $P$ in \texttt{ACE}}
\If{$P$ in layer \texttt{}}
    \State \texttt{protocols} $\gets \texttt{protocols} + P$
\EndIf
\EndFor
\State return \texttt{protocols}
\EndProcedure

\end{algorithmic}
\end{algorithm}

 \begin{table}[b!]
 \begin{center}

  \caption{Example of protocol merging between two devices}
  \label{tab:prot_matching}
  \begin{tabular}{cccccc}
    \toprule
    & Network & Transport &  Src Port & Dst Port \\
    \midrule
    \multirow{2}{*}{\rotatebox[origin=c]{90}{Dev1}}  
    &  IPv4 & UDP & any & any \\
     &  any & TCP & 5000 & any\\
    \midrule
    \multirow{2}{*}{\rotatebox[origin=c]{90}{Dev2}} 
     &  any & any & 5000 & 400\\
     &  IPv6 & any & any & 8080\\
     \midrule
     \midrule
     \multirow{3}{*}{\rotatebox[origin=c]{90}{Merged}} 
     &  IPv4 & UDP & 5000 & 400\\
     & IPv6 & TCP & 5000 & 8080\\
     & any & TCP &  5000 & 400\\
     
    \bottomrule
  \end{tabular}
  \end{center}
\end{table}

\begin{algorithm}[t!]
\caption{\acepruning}
\label{alg:protocol_pruning}
\begin{algorithmic}[1]
\State initialize $L, S, C, A_L, A_C$ to \texttt{null} 
\Procedure{Pruneacetree}{$PT$}
\For{each leaf $L$ in the Protocol Tree $PT$}
    \For{each $L$'s sibling $S$ in the Protocol Tree $PT$}
        \If{$L\subseteq C$}
            \State Prune($L$)
            \State continue to the next leaf $L$
        \EndIf
    \EndFor
    
    \For{each $L$'s cousin $C$ in the Protocol Tree $PT$}
        \If{$L \subseteq C$}
        \State $n \gets 1$
        \State $A_L \gets nthAncestor(L, n)$
        \State $A_C \gets nthAncestor(C, n)$
        \While{$A_L$ is not \texttt{null} and $A_L <> A_C$ }
            \If{$A_L\subset A_C$}                    
                \State Prune($L$)
                \State continue to the next leaf $L$
            \EndIf
            \State $n \gets n+1$
            \State $A_L \gets nthAncestor(L, n)$
            \State $A_C \gets nthAncestor(C, n)$
        \EndWhile
        \EndIf
    \EndFor
\EndFor
\EndProcedure
\Procedure{nthAncestor}{$Node$,$n$}
\State initialize $A_{Node}$ to \texttt{null} 
\State $c\gets 0$
\While{$c < n$} 
    \State $A_{Node}\gets Parent(Node)$
\State $c \gets c+1$
\EndWhile
\State return $A_{Node}$
\EndProcedure
\end{algorithmic}
\end{algorithm}

We illustrate an example of this process in Table \ref{tab:prot_matching}, where simple ACLs for two devices are presented (two ACEs per device). In this table, an intersection between the ACEs of these two devices exists. All possible pairs of ACEs from source and destination devices are checked against one another and the common factors are saved. 
The protocol stack in this example contains the Transports Layer and Network Layer. The first ACE of the first device is \texttt{[IPv4, UDP, any, any]}, representing the network, transport, source port, and destination port respectively. The first ACE of the second device is \texttt{[any, any, 5000, 400]}. By merging these two ACEs we find out that these two devices can only communicate if the network layer protocol is IPv4, the transport layer protocol is UDP,  and source and destination ports are 5000 and 400, respectively. After a comprehensive matching of all possible combinations of the ACEs from both devices, the result is three merged protocols. This result is presented in the 3rd row of Table \ref{tab:prot_matching}.

\begin{figure*}[t]
   \begin{center}
    \includegraphics[width=1\textwidth]{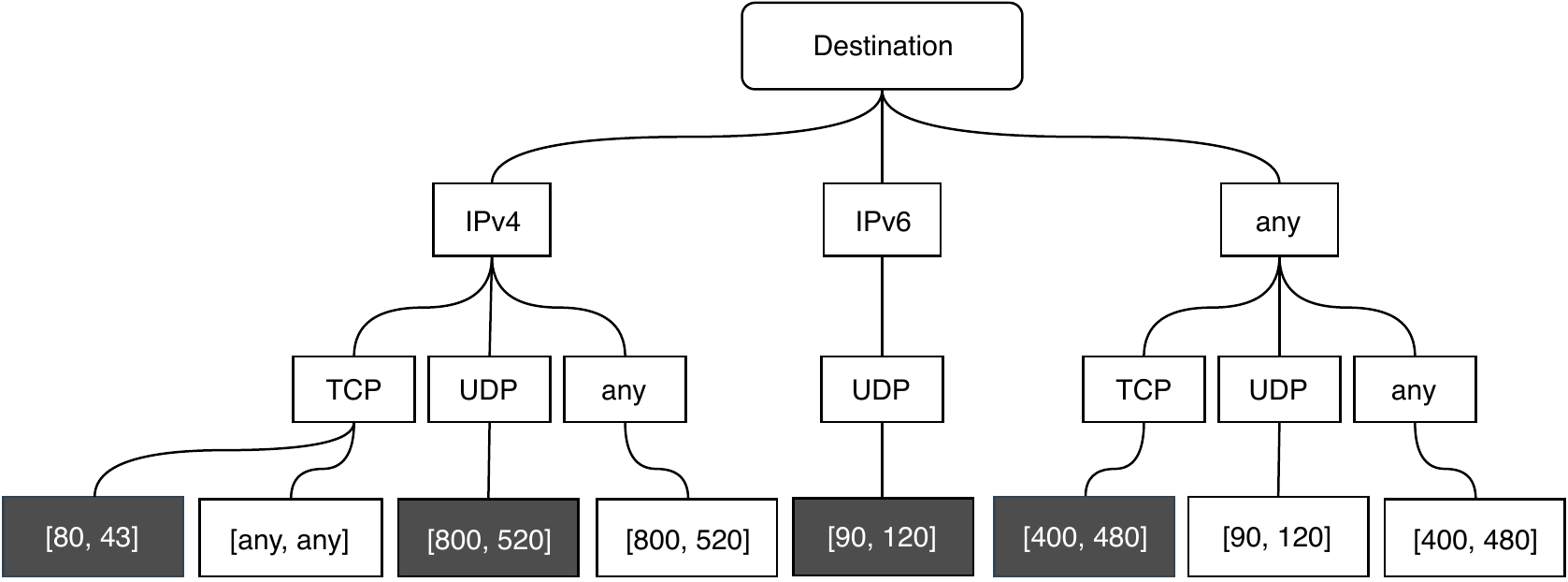}
    \caption{The tree structure containing the protocol information for a particular destination. The first children are the network protocols, the second children are the transport protocols and the leaves are pairs of source and destination ports. The nodes colored as dark gray could be removed as a super-set of them already exists in the tree. }
    \label{fig:pruning1}
    \end{center}
\end{figure*}

\subsection{\acetree{}} \label{sec-acetree}

When merging the protocols of the two \mudfile{}s that contain many ACEs, redundant protocols are a likely result. For example, suppose two ACEs are merged to \texttt{[IPv4, UDP, 400, 600]} and another two ACEs merged to \texttt{[IPv4, UDP, any, any]}. The result of the second merge in this example is a super-set of the first protocol, and therefore the first one should be pruned to prevent redundancy and further confusion.

To implement this efficiently for each IoT device, a tree structure was created and associated with each communication destination of that device. Each level of this tree contains information about a layer of the ACE protocol stack. Note that for each layer in TCP/IP model, one could add more than one level in the tree as we will present in our next example. Moreover, at each level, we have added a wildcard node in case the communication is allowed through multiple protocols in that particular layer. The recursive implementation of this procedure is presented in Algorithm \ref{alg:protocol_tree}.

\begin{table}[b!]
 \begin{center}
\caption{Result of the root to leaf tree traversal from the trees shown in Fig.  \ref{fig:pruning1} (Original) and Fig.  \ref{fig:pruning2} (Pruned)}
\label{tab:pruned_protocols}
\begin{tabular}{cccccc}
  \toprule
    & Network & Transport & Src Port & Dst Port \\
    \midrule
    \multirow{8}{*}{\rotatebox[origin=c]{90}{Original}}  
      & IPv4 & TCP & 80 & 43 \\
      & IPv4 & TCP & any & any \\
      & IPv4 & UDP & 800 & 520 \\
      & IPv4 & any & 800 & 520 \\
      & IPv6 & UDP & 90 & 120 \\
      & any & TCP & 400 & 480 \\
      & any & UDP & 90 & 120 \\
      & any & any & 400 & 480 \\
    \midrule
    \multirow{4}{*}{\rotatebox[origin=c]{90}{Pruned}}  
      & IPv4 & TCP & any & any \\
     & IPv4 & any & 800 & 520\\
     & any & UDP & 90 & 120\\
     & any & any & 400 & 480\\

    \bottomrule
  \end{tabular}
  \end{center}
\end{table}

As an example for Algorithm \ref{alg:protocol_tree}, we present a \acetree{} built from the set of ACEs presented in the \textit{Original} row in Table \ref{tab:pruned_protocols}. In this example, the protocol stack has simply two layers: Network Layer and Transport layer. However, as you can see, we have more than one level in the tree associated with the Transport layer, i.e. transmission protocols (TCP/UDP) and ports. 

In simple words, for each ACE, the algorithm does as follows: it starts from the lowest layer (in this case Transport layer), gets the protocols associated with that layer, and if they are more than one, it adds a wild card to that level of the tree. The \acetree{} containing the information of ACEs presented in the \textit{Original} row of Table \ref{tab:pruned_protocols} is presented in Fig. \ref{fig:pruning1}.

\subsubsection{Pruning \acetree{}} \label{sec-acepruning}

In this section, we describe how the \acetree{} is pruned. An example is provided for each of the scenarios where all the examples are based on the ACLs provided in \textit{Original} row in the Table \ref{tab:pruned_protocols} which are depicted as a \acetree{} in Fig.  \ref{fig:pruning1}. Consider the following notations: 
\begin{itemize}
    \item \texttt{$L$}: denoting one of the leaves of the \acetree{}
    \item \texttt{$S_i$}: denoting $ith$ sibling of the leaf \texttt{$L$}
    \item \texttt{$C_i$}: denoting $ith$ cousin of the leaf \texttt{$L$}
    \item \texttt{$A_{n}(L)$}: denoting $nth$ ancestor of the leaf \texttt{$L$}
    \item \texttt{$A_{n}(C_i)$}: denoting $nth$ ancestor of cousin node \texttt{$C_i$}
\end{itemize}

\begin{figure}[t!]
    \centering
    \includegraphics[width=0.5\linewidth]{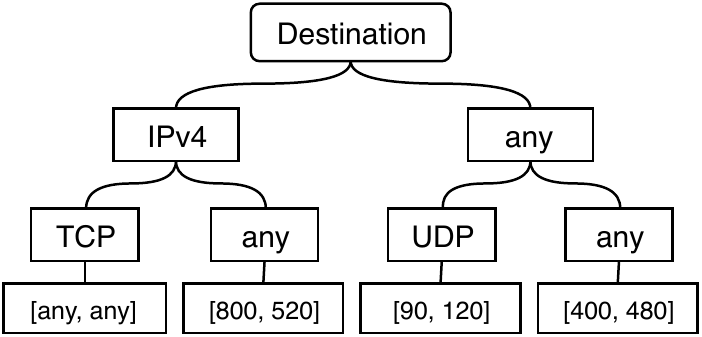}
    \caption{Result of the protocol pruning}
    \label{fig:pruning2}
\end{figure}

Each leaf node \texttt{$L$} in the \acetree{} can be pruned if it satisfies one of the following conditions: 
\begin{itemize}
    \item Consider \texttt{$L$} has a sibling \texttt{$S_i$} and \texttt{$L\subset \texttt{$S_i$}$}. In this case, \texttt{$L$} can be pruned with no futher conditions\\
    \textbf{Example}: Consider the leaf \texttt{[80,43]} and its sibling \texttt{[any,any]}. As can be seen, \texttt{[80,43]}$\subset$ \texttt{[any,any]}, hence \texttt{[80,43]} can be pruned. Note that the upwards tree traversal stops without any outcome when either we reach to the \texttt{root} or when \texttt{$A_{n}(L) = \texttt{$A_{n}(C_i)$}$}.

    \item Consider \texttt{$L$} has a cousin \texttt{$C_i$} where \texttt{$L\subseteq C_i$} and we traverse the tree starting from both \texttt{$L$} and \texttt{$C_i$} simultaneously up towards the root of the tree. \texttt{$L$} then could be pruned if at any point during traversal \texttt{$A_{n}(L)$} becomes a sibling of \texttt{$A_{n}(C_i)$} and  \texttt{$A_{n}(L)\subset \texttt{$A_{n}(C_i)$}$}. \\
    \textbf{Example}: Consider the fifth leaf of the tree (counting from left to right) \texttt{[90, 120]} and its sixth cousin \texttt{$C_6$} = \texttt{[90, 120]} which is the seventh leaf of the tree. As can be seen, \texttt{[90, 120]}$\subseteq$\texttt{[90, 120]} indicates that the first condition holds, i.e. \texttt{$L\subseteq C_i$}. As we traverse the tree towards the root by visiting the ancestors of each of the two target nodes, we see that \texttt{$A_{2}(L)$}, i.e. \texttt{IPv6} node, is a sibling of \texttt{$A_{2}(C_i)$}, i.e. \texttt{any} node, and \texttt{IPv6}$\subset\;$\texttt{any} indicating that \texttt{$A_{2}(L)\subset \texttt{$A_{2}(C_i)$}$}. Hence, the fifth leaf \texttt{[90, 120]} could be pruned. \\
    Another example in this case would be the third leaf \texttt{[800, 520]} and its third cousin \texttt{$C_3$}, i.e. the fourth leaf, \texttt{[800, 520]}. Since \texttt{[800, 520]}$\subseteq$\texttt{[800, 520]} and their first ancestors, i.e. parents, are siblings and \texttt{$A_{1}(L)\subset \texttt{$A_{1}(C_3)$}$}, therefore the third leaf \texttt{[800, 520]} can be pruned. We illustrate the pruned version of \acetree{} of Fig. \ref{fig:pruning1} in Fig. \ref{fig:pruning2}.

\end{itemize}

%-------------------------------------------------------------------------------
\section{Implementation}
\label{sec:implementations}
%-------------------------------------------------------------------------------

We implemented \toolname{} in JavaScript for two main reasons: the prevalent visualization libraries and enormous visualization capabilities in JavaScript, and the possibility of creating both a stand-alone application and a web application with minimum changes to the codebase. 

\begin{figure*}[b!]
  \centering
  \includegraphics[width=1\textwidth]{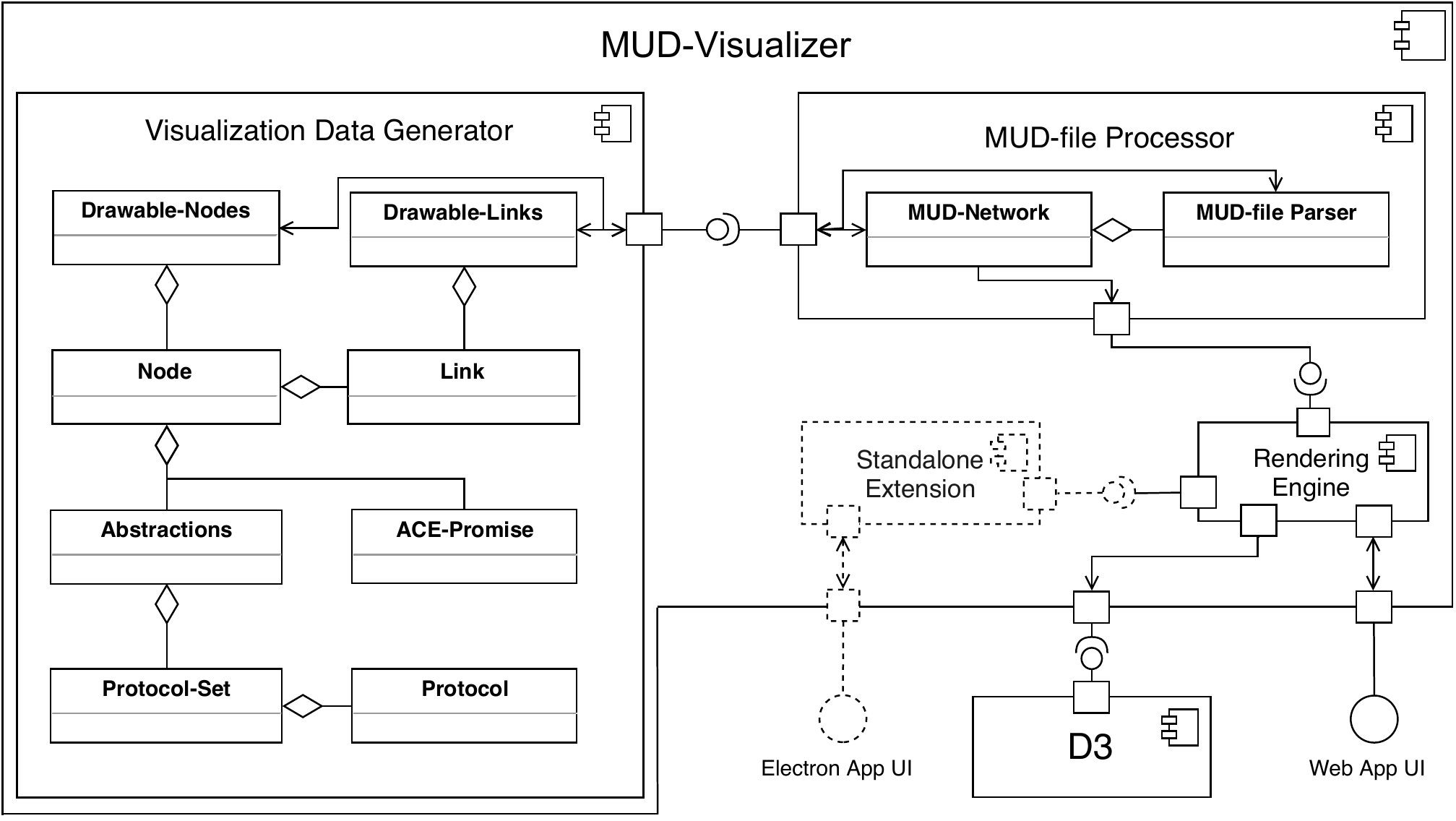}
  \caption{UML Diagram of the \toolname{}. The dotted line used for the Electron UI is merely for the two UIs, i.e. Web and Electron, to be distinguishable.}
  \label{fig:uml}
\end{figure*}

The UML diagram of the \toolname{} is presented in Fig.  \ref{fig:uml}. As shown, the D3 library\footnote{https://d3js.org/} is a vital component to visualizing the \mudfile{}s in \toolname. The main function of \toolname{} is to provide the appropriate data to the D3 library. There are four main internal components of the \toolname: \mudfile{} processor, visualization data generator, rendering engine, and standalone extension.

The \textbf{\mudfile{} Processor} initially parses the \mudfile{}s and extracts the data needed for the identification of possible flows. This information includes the MUD-URL, manufacturer, incoming and outgoing ACE, existence of my-controller or controller nodes, and associated data defined for the seven optional fields in Section~\ref{sec:abstractions}. Using this data the connection between the instances of the \mudfile{}s are analyzed. This includes whether or not two nodes should connect and if so, what are the assumptions about the protocols allowed between them. Note that the rules for each extension in Section \ref{sec:abstractions} might be different and are kept separately for further analysis in the \mudfile{} Processor component. In the case of a my-controller node, the required \textit{promises}\cite{friedman1976impact} are also saved so that they can be fulfilled by the user later on. Finally, the protocols are stored and merged in a way to minimize the redundant information, as was previously described in the sub-section entitled \nameref{sec-acemerging}.
% This is described in more details in the sub-section entitled \nameref{sec-acemerging}. 
Moreover, further information is requested from the user if needed, e.g., the configuration of my-controller nodes. This step of \toolname{} requires that the developer explicitly recognize their assumptions about the user or contextual knowledge required for configuration.

The \textbf{visualization data generator} converts the data extracted from \mudfile{}s into structures for the following visualization. These structures include nodes, links, and direction of the links as well as the corresponding protocols and rules that are assigned to the \mudfile{}s. For instance a simple \mudfile{} describing an IoT device with a domain-name abstraction that is only allowed to communicate with a remote server would generate several nodes and links including the IoT device and the remote server. In contrast, a device that should only communicate internally might seek to connect to a device that is constrained to only connect to a manufacturer. The concepts previously described in \nameref{sec:methods} section including building the \acetree{} in \ref{sec-acetree} and \acetree{} pruning in \ref{sec-acepruning} are both implemented in the \textit{Abstractions} module of this component.

The \textbf{rendering engine} is the component that combines all data generated by the other components and creates the final visualization. This component has several responsibilities, including interfacing with \mudfile{} Processor, the visualization libraries (i.e., D3), and also the Web App UI. If the application is running as a stand-alone app, it also communicates with the following component. 
    
The extension, called \textbf{Standalone Extension} is not used for the \toolname{} web app. It enables a stand-alone version of the \toolname. It consists of the components above, calls for those components, and the main script for the Electron UI application.

  \begin{figure*}[t!]
    \centering
    \includegraphics[width=1\linewidth]{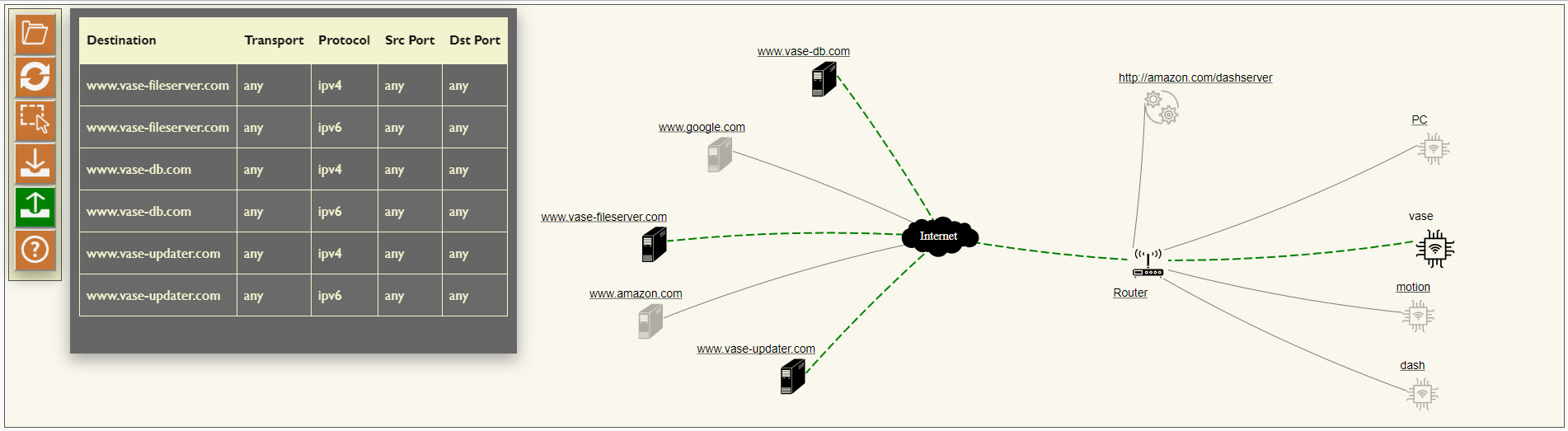}
    \caption{Screenshot of the UI of the \toolname}
    \label{fig:UI}
\end{figure*}

%-------------------------------------------------------------------------------
\section{Results}\label{sec:results}
%-------------------------------------------------------------------------------

\begin{figure*}[!t]
    \centering
    \includegraphics[width=1\linewidth]{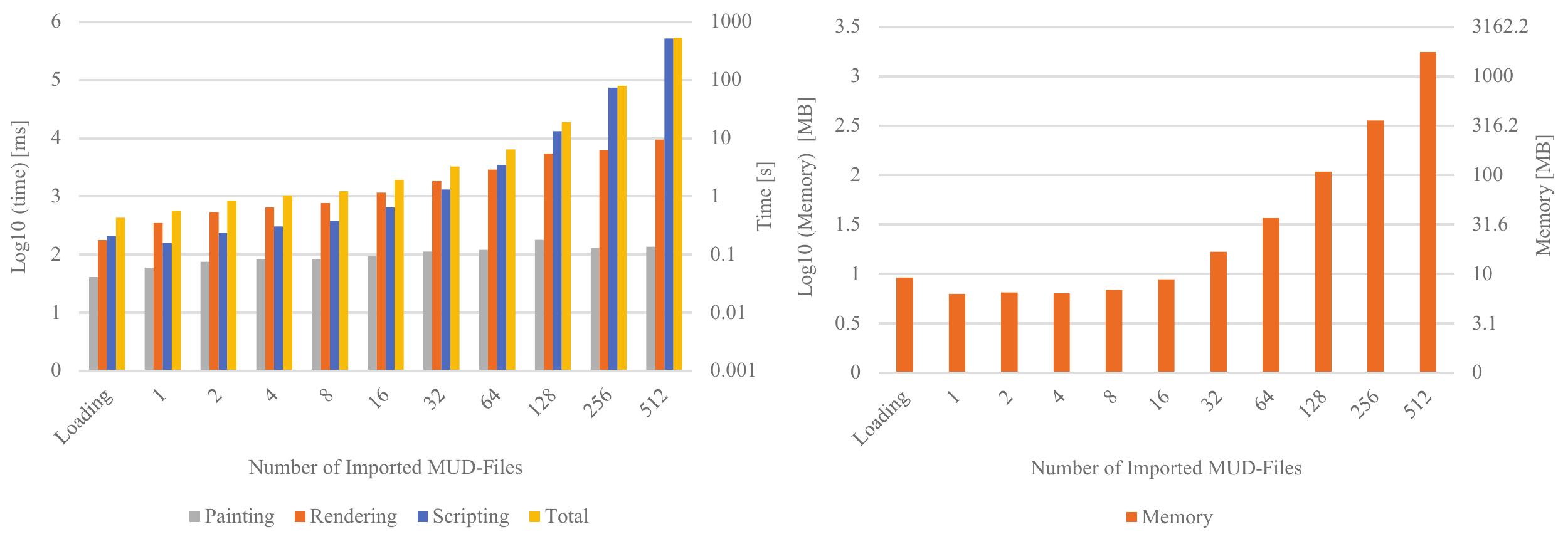}
    \caption{Performance evaluation of \toolname{} for first-time loading as well as importing 1 to 512 \mudfile{}s. Left: individual and total runtime for Painting, Rendering, and Scripting. Right: peak Javascript heap memory usage. Both charts are presented in logarithmic scale with the primary axis indicating the logarithmic values and secondary axis indicating actual values in \textit{seconds} and \textit{MegaBytes} respectively. The maximum runtime is for loading 512 \mudfile{} which is equal to 526 seconds and the corresponding memory usage is 1754 MB.} 
    \label{fig:benchmark}
\end{figure*}

The UI of the \toolname{} is shown in Fig. \ref{fig:UI}. This screenshot is consistent for both the stand-alone version and web app version of the tool. 

Given that the Electron framework also supports the DevTools, we used the Chrome Performance Analysis tool available as part of the Devtools for benchmarking the web application version of the \toolname{}. We considered the time spent for \textit{Scripting}, \textit{Rendering}, and \textit{Painting} as well as peak Javascript heap memory usage. Our experiments were run on a Macbook Pro late 2013 computer with 2.6GHz Quad-Core Intel Core i7, 16GB 1600 MHz DDR3 RAM, and $2880x1800$ pixels of screen space. 

The benchmark results are presented in Fig. \ref{fig:benchmark}. We evaluated the performance of \toolname{} when loaded for the first time, i.e. indicated as \textit{Loading} in the figure, as well as when we import \mudfile{}s. The number of \mudfile{}s that was used in the benchmark ranged from 1 to 512. To evaluate the scalability of \toolname{}, we used copies of a relatively heavy \mudfile{} created by \textit{mudmaker} which includes five out of seven implemented abstractions, i.e. all abstractions except \textit{my-controller} and \textit{model}. Please recall that \textit{my-controller} requires end-user to manually enter data about their selected point of control. Had we included \textit{my-controller} the results would have been dominated by user response time. Examining user interaction is a component of our future work. Also, the \textit{model} abstraction would result in all copies of the sample \mudfile{} to communicate with each other in a local network. Therefore, we decided not to include that to make the benchmarking more rigorous by letting the \toolname{} process all other abstractions.

Note that in a real-world setting, although the enterprises might have thousands of IoT devices in their networks, the \textit{type} of devices in their network is significantly lower than that. For instance, a hospital that has 2k MUD-compliant smart bulbs in its network will have bulbs of a few brands and types. In this case, if a hospital system administrator tries to use \toolname, they will not need to load 2k MUD-files but rather a handful of them. In other words, our benchmark shows the scalability of \toolname{} with regard to the \textit{type} of \mudfile{}s not the number of MUD-compliant devices in the network.  The threat model and risk posture of enterprise will determine if new devices should be manually added, if \toolname{} should interact with automated detection and identification, or if there should be some combination of these strategies. 

The runtime benchmark data in Fig. \ref{fig:benchmark} indicates that the total time gets very close to scripting time as the number of \mudfile{}s increases. This is because for each \mudfile{} that is newly imported, its relation and interaction with other \mudfile{}s with respect to all MUD abstractions should be analyzed and processed. The maximum loading time is for 512 \mudfile{}s which is slightly longer than 10 minutes and the memory that the application needs exceeds slightly higher than 1754 MB. Be advised that this benchmark is performed rigorously and even enterprise networks barely have 512 \textit{types} of \mudfile{}s in their network each being as heavy as the one we used in this benchmark. Moreover, verification of \mudfile{} using the \toolname{} is not performed on short intervals and is done only when, for instance, a new device is introduced.

%-------------------------------------------------------------------------------
\section{Conclusions and Future Work}
%-------------------------------------------------------------------------------
% future: controller mud files, config of the network, DHCP 
In this work, we described a tool for visualizing the interaction of \mudfile{}s which also explicitly identifies any information required by the use of the controller options. The challenge in visualizing the MUD-files is the way they interact with each other and how the ACL of the devices affects the communication between them. We presented methods for addressing this challenge and implemented \toolname{} and made it open-source and publicly available on GitHub. The main purpose of \toolname{} is to facilitate the review and validation phase of MUD deployment for developers, engineers, and system administrators. We also performed a benchmark for runtime and memory consumption of the tool and showed that it can be used on a personal computer to load hundreds of \mudfile{} for evaluation. 

Our future work includes several phases. First, a user study in which we examine the practicality of the \toolname{} and the extent to which it can help the target audience. Second, there are a few points in the tool that we are particularly interested to improve, including visualizing the controllers' \mudfile{}, support for including network configuration in the visualization, e.g., the IP address of IoT devices and the corresponding controllers, DHCP configuration, etc. Third, we want to test the functionality that facilitates the local modification of the \mudfile{}s without creating the opportunity for an attacker to move around these. Essentially our goal is to allow decreased but not increased connectivity. Finally, we want to implement the abstraction analysis of \toolname{} in parallel to improve the performance and scalability even more. 

%one of our other goals it to use automatic creation of mud files in a laboratory setting and determine how these work with MUD visualizer. DK this would be a good paper for you!

% \toolname is made available both as a stand-alone tool and as a web application.
% \toolname is open-source and publicly available in GitHub. 

% The major next steps in this study includes adding the modification capability to the problem so that the user would be able to modify the MUD-files during visualization.

% \toolname is open-source and publicly available in github under the name MUD-Visualizer\footnote{https://github.com/iot-onboarding/mud-visualizer}. The web app version of MUD-Visualizer is also integrated in the mudmaker website\footnote{https://www.mudmaker.org/mudvisualizer.html}. 

%-------------------------------------------------------------------------------
\section*{Availability}
%-------------------------------------------------------------------------------

\toolname{} is made publicly available in GitHub at \href{https://github.com/iot-onboarding/mud-visualizer}{https://github.com/iot-onboarding/mud-visualizer}
% \textit{anonymized}  % anonymized
and can be used both as a stand-alone tool and as a tool integrated into web apps. 
% As an example it is currently available and integrated in \textit{mudmaker} website\footnote{https://www.mudmaker.org/mudvisualizer.php}.

%\section{Acknowledgments}
%-------------------------------------------------------------------------------
\section*{Acknowledgments}
%-------------------------------------------------------------------------------

% This research was supported by Cisco Research Award funding. This material is based upon work supported by the National Science Foundation under Grant CNS 1814518 and CNS 1565375. Any opinions, findings, and conclusions or recommendations expressed in this material are those of the author(s) and do not necessarily reflect the views of the NSF.
This research was supported in part by the National Science Foundation awards CNS 1565375 and CNS 1814518, as well as the grant \#H8230-19-1-0310, Cisco Research Support, Google Research, and the Comcast Innovation Fund. Any opinions, findings, and conclusions, or recommendations expressed in this material are those of the author(s) and do not necessarily reflect the views of the National Science Foundation, Cisco, Comcast, Google, nor Indiana University.

%-------------------------------------------------------------------------------
\bibliographystyle{splncs04}
\bibliography{refs}

\end{document}